\documentstyle[12pt]{article}
\begin{document}

{\Large\bf PARTICLE DIFFRACTION}\\

\vspace*{0.3cm}
{\Large\bf AT HIGH ENERGIES}\\

\vspace*{1cm}
{\large\bf Vladimir A.Petrov}\\

\vspace*{0.5cm}
\centerline{{\it Institute for High Energy Physics, 142284, Protvino, Russia}}

\vspace*{1cm}
Diffraction of light was described by Italian physicist 
Grimaldi in his book published  in 1665 . 
One of the first (and wrong) explanations was given by Newton, who also
contributed a lot into the experimental 
discovery  and the study of new diffractive phenomena. Newton's
explanation of light diffraction was based on a corpuscular theory of light.
However, in the beginning of the XIX century the famous ``Poisson's puzzle''
(the prediction of a light spot in the center of the geometric shadow, a
consequence of the Fresnel's wave theory of light) and its experimental
confirmation affirmed  wave nature of light for hundred years, until
Einstein and Stark disovered that light demonstrated particle properties
as well.

From the observation of the diffractive pattern one can judge about the size
and the shape of the scatterer. At present this field is a highly developed branch of
applied optics, with innumerous uses and applications in technology.

Since a fundamental guess made in~1923 by Louis de Broglie on wave properties of
matter, confirmed experimentally by Stern in Germany and by Davisson 
and Germer in the USA, this peculiar quantum behaviour has found a lot of
applications. The main lesson was that undulatory or corpuscular properties are
inherent to all natural phenomena, though  one or another 
aspect may dominate dependent on conditions. 

High energy physics is usually synonymous  to ``particle
physics''. New phenomena in this field are related either to the discovery of new
particles or to some typical particle --- like effects as, say, Bjorken scaling in
deeply inelastic scattering,  or high $p_\bot$ jets, or else. In space-time
language these regimes mean the probe of small distances.

However there is  a field in high energy physics which even at very high 
energies is not related to short distances but rather to large (at nuclear
scales) distances. Such are phenomena like small angle hadron  scattering (elastic or
inclusive). It is well known feature of these processes that the angular
probability distribution of the scattered particle shows a typical diffractive
pattern with a maximum at zero angle followed by the dip and, in some cases,
second maximum.\footnote{Interesting discussion of ``high energy diffraction''
is contained in Ref.[1].} Here we deal with wave properties of hadrons.

From such a distribution one can conclude about the size of the scatterer, or,
more properly, the ``interaction region''.

An interesting feature of these ``size measurements'' is that the size appears
to be energy dependent. This would correspond to dependence of the visible size
of a lit object on the frequency (or the wavelength) of falling light. 

Modern theory limits this energy dependence of the transverse (w.r.t. the incident
beam $(s)$) size by a ``maximal radius'', $R_0\approx (1/m_\pi)\log E$, 
where $m_\pi$ is the pion mass 
($1/m_\pi$ is the famous Yukawa radius), and $E$ is the center-of-mass
energy. Logarithmic dependence of the strong interaction transverse range was
derived by W.~Heisenberg in the framework of some model 
of high-energy collisions as early as in~1952.
Later M.~Froissart obtained the same limit on more general grounds in~1961,
and, finally, A.~Martin gave in~1966 a rigorous proof based on the first
principles of quantum field theory. Lower bounds on the strong interaction
radius were given by A.~Logunov and Nguen Van Hieu~[2].

Experiments confirm the energy dependence of the transverse interaction range
which weakly grows with energy (but is far below the Heisenberg-Froissart-Martin radius
$R_0$). 

Whereas one can extract the transverse interaction radius from the differential
cross section, 
what can one say about the longitudinal size of the interaction region 
or the interaction time?

Theoretically, the problem was addressed in an early paper by  Wigner in the
framework of non-relativistic quantum mechanics~[3]. One can also mention
papers~[4]. In these papers the longitudinal range was related to some
derivatives of the phase of the scattering amplitude. Unfortunately this
procedure needed the knowledge of the off-mass-shell amplitude.

A different approach was used in Ref.~[5], where the effective longitudinal
size was estimated to grow with energy as fast as $E/m^2$. This is very
interesting because at energies of the future Large Hadron Collider the
longitudinal interaction range can achieve atomic scales. 

Unfortunately at present  no  way to extract this size from the
measured characteristics is known. Some hopes refer to nuclear targets where
more than one nucleons could be involved into interaction with a 
``long''projectile. 

If one imagines that the size and the shape of the interaction region are extracted
from a complete enough set of experimental data, then the problem is to
understand the information obtained on the basis of present theoretical
frameworks. Let us consider a high energy collision in the laboratory frame when one
hadron (nucleon, or nucleus in practice) is at rest (``observer'') while
another one flies on. Energy dependence then may be mainly related to the
projectile, which seems to be longer in the longitudinal direction and larger
in the transverse ones. 

Is not it in an apparent contradiction with the special relativity which
predicts that the longitudinal size should \underline{decrease} with the  growing velocity while
the transverse ones remain intact? In fact there is no contradiction. The
matter is that a particle is a quantum object which is hardly a rigid sphere as
one could imagine in  a classical manner. This is a quantum system which
fluctuates into various virtual states which have their own lifetimes and
sizes. The latters are by no means Lorentz invariant. Moreover, the maximal
radius, $R_0$, refers, in the transverse plane, to 
distances between the points taken at 
different times, and this is
not the same as the ``instantaneous size'' of special relativity.

Quantum fluctiations have specific features which should be related to modern
views of microstructure of particles. 
For strongly interacting particles this is quantum chromodynamics,
or shortly, QCD.

QCD gave many insights into understanding of phenomena, related to short
distances (``hard processes'').

Unfortunately, QCD is still not very effective when applied to large distance
(``soft'' or diffractive) processes. In the framework of Regge approach these
are some attempts to obtain the leading Regge trajectory perturbatively. In
spite of some progress serious problems remain to be resolved. 
One of these problems is that the method of quantum perturbations, which works
nicely at short distances, fails at large distances. This is related to the
confinement problem, i.e. absence of quarks and gluons in asymptotic states
detected by the measuring apparatus. 

It may well happen that ``particle'' approach, where quarks and gluons take
part in the process of scattering as constituents of colliding hadrons, is not
relevant to diffractive phenomena, which are more adequate to  wave aspects.
In this case it could be more appropriate to study some (gluon) field
configurations which are beyond  reach of usual perturbative treatment. 
That is why projects like TOTEM at LHC should be considered not just as an
inevitable price for a precise measurement of luminosity but rather as a unique
source of information about sizes and shape of the hadron interaction region.
Explanation and description of these is a formidable task for QCD. 

As a conclusion I should like to stress again that the experimental study of
diffractive hadron scattering is important and interesting because:

1. Energy-dependent shape of the interaction region 
is interesting both from general quantum and relativistic points of view;

2. The interpretation of data can promote the new 
development of QCD at large space-time scales. This is definitely related to
the long-standing confinement problem, which, as we see, is important not only
at low energies. 

\vspace*{0.4cm}
{\bf References}\\
{}[1] A.P.~Samokhin. In Proc. XVIII Workshop on 
High Energy Physics and Field Theory
(Protvino, June 26-30, 1995). Ed.~V.A.Petrov et al. Protvino 1996.\\
{}[2] A.A.~Logunov and Nguen Van Hieu. Teor.Mat.Fiz., v.1 (1969)~375.\\
{}[3] E.~Wigner. Phys.Rev., v.98 (1955)~98.\\
{}[4] M.~Froissart, M.~Goldberger and K.~Watson. Phys.Rev., v.131 (1963)~2820.\\
{}[5] V.N.~Gribov, B.L.~Ioffe and I.Ya.~Pomeranchuk. Sov.Journ.Nucl.Phys., v.2
(1967)~768. 
\end{document}